\documentstyle[aps,preprint]{revtex}

\tightenlines

\begin{document}

\draft 


 \widetext

\title{Effect of Quantised Lattice Fluctuations on the
Electronic States of Polyenes}

\author{William Barford$^{1*}$, Robert J. Bursill$^2$ and 
Mikhail Yu 
Lavrentiev$^{1**}$}

\address{
$^1$Department of Physics and Astronomy, The University of 
Sheffield,
\\
Sheffield,  S3 7RH, United Kingdom.
\\
$^2$Department of Physics, UMIST, Manchester,  M60 1QD, United 
Kingdom.
}

 \maketitle

\mediumtext

\begin{abstract}

We solve a model of interacting electrons coupled to longitudinal phonons using the density matrix renormalisation group method.
The model is parametrised
for polyenes.
 We calculate the ground state,
and first excited odd-parity singlet and triplet states; and we
investigate their energies, and
bond length changes and fluctuations for up to $30$ sites.
The transition energy and the soliton width of the triplet state show
 deviations from the adiabatic approximation
for chain lengths larger than the classical soliton size,
 because of de-pinning by
the quantised lattice 
fluctuations. 

\end{abstract}



The 
inter-play of electron-electron interactions and electron-lattice 
coupling in polyene 
oligomers and {\em trans}-polyacetylene, (CH)$_x$,
 results in a rich variety of  low energy excitations.  
These excitations include triplet states of soliton-antisoliton pairs,
singlet states comprising bound pairs of triplets, and exciton-polarons.
Within the adiabatic (or classical) 
approximation\cite{footnote1} the 
nature and energy of these excitations is now fairly well 
understood.  A  parametrised 
Pariser-Parr-Pople-Peierls model, solved within the adiabatic
approximation, predicts  accurate 
excitation energies for oligomers of up to $20$ or so sites 
\cite{bursill99} \cite{barford00}.   However, for longer chains there are 
deviations from the polyacetylene thin film results.  
  These discrepencies are partly explained by  the  self-trapping of the 
excited states by the lattice\cite{footnote2}.
The calculated energies deviate from a linear extrapolation in $1/N$ as the
chain length becomes larger than the solitonic structures.  
We can use this deviation in the energy to estimate an upper
bound for the self-trapping energy.
This deviation is $0.4$ eV for the optically allowed ($1^1B_u^-$)
 state, $0.7$ eV for the even-parity singlet 
($2^1A_g^+$) state and $0.3$ eV for the lowest lying triplet 
($1^3B_u^+$)\cite{bursill99}.
Furthermore, a linear extrapolation in $1/N$ of the 
short oligomer experimental values
 predicts infinite chain energies of the 
 $1^1B_u^-$ and $2^1A_g^+$ states close to those observed in
polyacetylene thin films\cite{kohler}, suggesting that self-trapping
may be a partial artefact of the adiabatic approximation.
Thus, the question remains as to the role of quantised
lattice fluctuations,
 both on  the dimerisation of the 
ground state, and to the de-pinning of the excited states.
These fluctuations are the subject of this paper.

There have been a number of studies of quantised lattice 
dynamics in the ground state of the uncorrelated 
Su-Schrieffer-Heeger model \cite{su82}, 
indicating that fluctuations in the 
bond length are comparable to the bond length changes, but that the Peierls 
dimerisation is stable against such fluctuations.
There has also been a variational Monte Carlo study of an 
interacting electron-phonon model \cite{alder97}. 
However, there have been no studies of excited states, as the incorporation 
of quantised lattice dynamics into the correlated 
Pariser-Parr-Pople-Peierls model presents a formidable challenge.  

The advent of the density matrix renormalisation
group (DMRG)
method \cite{white92},\cite{book}
has enabled definitive model studies of correlated electron systems,
 including long
range interactions\cite{yaron98} and dynamical phonons
\cite{holstein}, \cite{caron97}. In this work we report the results of
extensive calculations on a model system which for the first time
affords us insight into the effect of quantised lattice dynamics
on the properties of excited states of long polyenes. Electrons,
interacting via long-range Coulomb forces, are coupled
to longitudinal phonons. The key results of this
calculation are that the de-pinning of excited states due to quantum
lattice fluctuations can become substantial as the conjugation
increases. In particular, there is
 a marked reduction in the energy
and increase in the soliton width of the triplet excited state.

The Hamiltonian, with free boundary conditions, is defined as \cite{caron97}
\cite{bursill99},
\begin{eqnarray}
{\cal H}
 & = & \hbar \omega \sum_{i= 2}^{N-1}
 \left( b_{i}^{\dagger} b_{i} + \frac{1}{2} \right) 
+ \hbar \omega_0 
\left( b_1^{\dagger} b_1  + b_N^{\dagger} b_N + 1 \right)
- \hbar \omega \sum_{i= 1}^{N-1}  B_{i + 1} B_{i}
+ 2 \Gamma t g  \sum_{i = 1}^{N-1}  
\left(  B_{i + 1} - B_{i} \right),
\nonumber
\\
& &  + \;
U \sum_{i =1}^N \left(n_{i\uparrow} -
\frac{1} {2} \right)
\left(n_{i\downarrow} -  \frac{1} {2} \right)
+ \frac{1}{2} \sum_{i  \ne j}^N  
V_{ij} (n_{i} - 1)(n_j - 1)
\nonumber
\\
& &  - \;
 t \sum_{i = 1,\sigma}^{N-1}
 \left( 1 +  g \left(B_{i +1} - B_{i}\right) \right) 
\left(c_{i+1 \sigma}^{\dagger} c_{i \sigma} 
+ c_{i \sigma}^{\dagger} c_{i+1 \sigma} \right).
\end{eqnarray}
$b_{i}^{\dagger}$ ($b_{i}$) creates (destroys) a phonon and 
$c_{i \sigma}^{\dagger}$ ($c_{i \sigma}$) creates (destroys) an electron 
on site $i$.
$B_{i} = (b_{i}^{\dagger} + b_{i})/2$, 
$g=(\lambda \pi 
\hbar \omega/ 2 t)^{1/2}$ 
and $\omega = \sqrt{2}\omega_0 = 
\sqrt{2 K /m}$.
We use the Ohno function for the Coulomb interaction:
$
V_{ij} = U / \sqrt{ 1 + (U r_{ij}/14.397)^2 },
$
where
the bond lengths are in \AA. The single and double 
bond lengths used in the evaluation of $V_{ij}$ are $1.46 \AA$ 
and $1.35 \AA$, 
respectively, and the bond angle is $120^0$. $t = 2.539$ eV, $U=10.06$ eV, 
$\Gamma = 0.602$, 
$\lambda =0.115$\cite{bursill99} and 
$\hbar \omega_0 = 0.2$ eV\cite{schugerl81}.    

The essential approach we adopt is an extension of the local 
Hilbert space reduction of \cite{zhang98} for a representative 
repeat unit, namely two lattice sites. Once a repeat unit Hilbert 
space is optimised it is then augmented with the system block in 
the standard finite lattice algorithm \cite{white92}.  Since the 
classical lattice geometry of excited states changes as the chain 
length increases, there is no {\em a priori} reason to suppose that 
the optimal repeat unit electron-phonon basis for the shortest 
chain is appropriate  for longer chains.  Thus, it is generally 
necessary to perform {\em in situ} optimisation, i.e.\ a repeat 
unit  Hilbert space is re-optimised when it forms part of the 
target chain size.  Generally, we expect {\em in situ} optimisation to 
be necessary whenever the short scale properties are modified by the
long scale properties.

We 
now outline the procedure in  more detail.
We begin with a six site lattice, composed of three repeat units, 
and optimise the repeat unit electron-phonon Hilbert space.  This is 
done by retaining the optimised states and 
`folding-in' some `bare' electron-phonon states (typically $16$).
Once the full
electron-phonon basis has been swept through for a particular repeat 
unit, the same procedure is applied to the next repeat unit until 
convergence is achieved.  Next, two  repeat units are augmented 
to form a four site block for the next chain size, i.e.\ $10$ sites.
The optimised basis for each repeat unit is retained.  
For $10$ sites and greater, each repeat unit is re-optimised 
(including the end units) by sweeping through the 
electron-phonon basis on the first finite lattice sweep.  
({\em In situ} optimisations in subsequent finite lattice sweeps
 were found to be 
unnecessary.)  During the {\em in situ} 
optimisation only a few states (typically $\le 50$) are 
retained for the environment blocks.  After completing the sweep
the optimised states of the repeat unit are retained for 
augmentation with the left hand block. 
Typically, $150$ states are used for the system and 
environment blocks during augmentation.

A key goal of this work is to study excited states, which we do 
by exploiting the particle-hole (\^J) and spin-flip (\^P) 
symmetries 
 of Eq.\ (1) \cite{footnote3}.  
The inversion symmetry is measured at the middle
of a finite lattice sweep.  
We 
have checked that setting $J=+1$ and $P=+1$ targets the 
$1^1A_g^+$ 
state, setting $J=-1$ and $P=+1$ 
targets the $1^1B_u^-$ state, and setting 
$J=+1$ and $P=-1$ targets the $1^3B_u^+$ state.

We now turn to the convergence tests.   We first establish 
convergence with respect to the number of optimised states per 
repeat unit.  Table I shows the ground state and the $1^1B_u^-$ 
transition energies for the six site chain for a maximum number of
 two and three bare phonons per site.
  We see that with $64$ states the transition energy has 
converged to within $0.001$ eV.   Next, we consider 
the ground state and the $1^1B_u^-$ transition energies as a 
function of the maximum
number of bare phonons per site, as shown in 
table II.   We see that the transition energy has essentially converged to 
$0.001$ eV with five phonons per site, and to within $0.02$ eV
 with two phonons  (which is  better than
experimental accuracy).  The converged  
$1^1B_u^-$ excitation energy of $4.62$ eV is very close to the 
classical result of $4.65$ eV. Notice also, that the average 
phonon occupation per site is ca.\ $0.2$.  Finally, we consider 
the convergence with super block Hilbert space size for $18$ sites.  
The convergence of the ground state energy is reasonable for up 
to $180 000$ states, and the transition energy has converged to 
better than $0.01$ eV.

Next, we consider in what way quantising the lattice degrees of freedom
leads to deviations from the adiabatic approximation.
We calculate the classical phonon 
displacement, 
$q_i = ( \hbar \omega /K)^{1/2} \langle B_i \rangle$
\cite{ehrendfreund87},
the bond length distortion
and the root mean 
square fluctuations in the bond length.
Fig.\ 1 shows the staggered bond length changes in the ground 
state of an 18 site chain for up to three phonons per site.  Also 
shown is the classical result\cite{bursill99}.  In the middle of the chain the 
phonon calculation is close to the classical result of a bond 
length distortion of  ca.\ $0.05 \AA$.  However, towards the end 
of the chain the phonon calculation predicts a somewhat larger 
distortion.  We find that in the limit of long chains the relative 
bond length distortion is ca.\ 0.9,  almost independent of the 
number of phonons per site, and close to previous 
theoretical\cite{su82} and experimental \cite{mckenzie92}
estimates.  We
note that the bond length fluctuations do not 
reduce the average bond length change for the linear polyenes 
considered here.  This is because the dimerised ground state 
 is not degenerate with respect to the state with the bond 
lengths reversed (in contrast to cyclic polyenes) 
\cite{barford00}, and thus there is no quantum mechanical 
tunnelling between the two dimerised states.

Lattice fluctuations do, however, de-pin the self-trapped solitonic 
structures of the excited states.  Fig.\ 2 shows the triplet state
transition 
energy as a function of inverse chain length for up to three
phonons per site, and for the classical approximation.  The 
energy in the classical approximation deviates markedly from a 
$1/N$ behaviour for chain lengths greater than $20$ sites, as 
shown in Fig.\ 2 and especially in \cite{bursill99} \cite{barford00}.  
We intrepret this as a result of the electronic wavefunction being 
trapped by the lattice structure.  The phonon calculations have essentially
converged by three phonons per site.  In constrast to the classical result, 
the converged phonon 
calculation shows a much weaker deviation from $1/N$ 
behaviour, leading to an expected correction of a few tenths of 
an eV in the infinite chain limit.                      

The solitonic structure of the triplet state supports this 
de-pinning hypothesis.  Fig.\ 3 (a) and (b) show the soliton 
structures for $18$ and $30$ sites, respectively.  The position of 
the defect, at roughly
the 4th.\ bond from the center, is roughly the same 
for   both chain lengths, and for both the classical and phonon 
calculations.   We expect this, as  its position is determined by 
the electronic component of the wavefunction \cite{barford00}.  
However, we can see that for the classical calculation the soliton 
width is virtually the same for both chain lengths (see also Fig.\ 
4 of \cite{bursill99}), whereas for the phonon calculation the 
soliton width is greater in the longer chain.
This indicates that the coupled electronic and lattice fluctuations
 lead
to an increased delocalisation of the wavefunction, and 
hence to a lower energy.
The $2^1A_g^+$ state is a bound state of two triplets, and it too is 
self-trapped
by the lattice.  Although we cannot target this state in 
the current calculation, the above discussion indicates that it will 
also be de-pinned by lattice fluctuations.

Finally, we consider the optically allowed excitonic 
($1^1B_u^-$) state.  According to the adiabatic approximation 
\cite{grabowski85},  this state creates a shallow 
polaronic distortion of the lattice, with self-trapping only 
becoming important for chain lengths longer than ca.\ $40$ 
sites\cite{bursill99}.  
Thus, we would not expect lattice fluctuations to play a 
significant role for the shorter chains which we have so far 
considered.  This is confirmed by the excitation energies shown 
in Table IV, indicating that the quantised lattice calculated 
energies are within $0.1$ eV from the classical result, and Fig.\ 
4, showing that the quantum and classical polaronic structures 
are virtually identical.

In conclusion, an extended DMRG method has been  applied to an 
interacting electron-phonon model of polyenes.    Quantum 
lattice fluctuations are shown to play an important role in the
de-pinning of the self-trapped excited states, leading to
corrections to the adiabatic approximation, and to an
expected reduction
of the transition energies of a few tenths of an eV for long chains.
Thus,  a full quantum
mechanical treatment of the Pariser-Parr-Pople-Peierls model
 gives remarkably accurate predictions for the
excited state energies of polyenes.

M. Yu.\ L. was supported by the EPSRC (U.K.) (GR/K86343).

\begin{table}[h]
\caption{
The ground state ($1^1A_g^+$) and $1^1B_u^-$ transition
energies (eV) as a function of the number of optimised states per repeat
unit  for the six site chain.
}
\begin{tabular}{cccc}
Bare phonons per site   &  Optimised states  & $E(1^1A_g^+)$ & 
$E(1^1B_u^-)-E(1^1A_g^+)$   \\
\hline
$2$ & $48$ & $-29.0539$ & $4.6138$ \\
$2$ & $64$ & $-29.0578$ & $4.6087$ \\
$2$ & $80$ & $-29.0588$ & $4.6087$ \\
$2$ & $96$ & $-29.0596$ & $4.6085$ \\
$2$ & $144$ (Exact) & $-29.0597$ & $4.6083$ \\
\hline
$3$ & $48$ & $-29.0687$ & $4.6229$ \\
$3$ & $64$ & $-29.0756$ & $4.6184$ \\
$3$ & $80$ & $-29.0772$ & $4.6182$ \\
$3$ & $96$ & $-29.0787$ & $4.6178$ \\
$3$ & $144$ & $-29.0793$ & $4.6173$ \\
\hline
\end{tabular}
\label{I}
\end{table}

\begin{table}[h]
\caption{
The ground state and  $1^1B_u^-$ transition energies (eV), 
and the average ground state phonon occupation per site 
for the six site 
chain.  There are $64$ optimised  states per repeat 
unit. 
}
\begin{tabular}{cccc}
Bare phonons per site   &  $E(1^1A_g^+)$  &
 $E(1^1B_u^-)-E(1^1A_g^+)$ & 
$\frac{1}{6} \sum_{i=1}^6 \langle b_i^{\dagger} b_i
\rangle_{1^1A_g^+} $  \\
\hline
$0$ & $-28.6681$ & $4.353$ & $0$ \\
$1$ & $-28.9992$ & $4.577$ & $0.090$ \\
$2$ & $-29.0578$ & $4.609$ & $0.137$ \\
$3$ & $-29.0756$ & $4.618$ & $0.164$ \\
$4$ & $-29.0810$ & $4.622$ & $0.185$ \\
$5$ & $-29.0832$ & $4.623$ & $0.192$ \\
\hline
Classical{\protect\cite{bursill99}} &--- & $4.646$ & --- \\
\hline
\end{tabular}
\label{II}
\end{table}

\begin{table}[h]
\caption{
The ground state and  $1^1B_u^-$ transition energies (eV) as 
a function of the number of system block states ($m$) and 
superbock Hilbert space size (SBHSS) for the $18$ site chain 
with two bare phonons per site.  There are $60$ optimised states 
per repeat unit.
}
\begin{tabular}{cccc}
$m$  & SBHSS & $E(1^1A_g^+)$ & 
$E(1^1B_u^-)-E(1^1A_g^+)$  \\
\hline
$125$ & $97700$ & $-89.3167$ & $3.141$  \\
$154$ & $140248$ & $-89.3259$ & $3.145$ \\
$173$ & $184288$ & $-89.3304$ & $3.147$ \\
\hline
\end{tabular}
\label{III}
\end{table}

\begin{table}[h]
\caption{
The  $1^1B_u^-$ transition energies
(eV) as a function of the 
number of sites, $N$.
}
\begin{tabular}{c|cccc|c}
$N$ & \multicolumn{4}{c|}
{Maximum number of phonons per site}& 
Classical {\protect \cite{bursill99}}\\
\hline
    & $0$ & $1$ & $2$ & $3$ &  \\
\hline
$10$ & $3.466$ & $3.758$ & $3.837$ & $3.864$ & $3.812$ \\
$14$ & $2.984$ & $3.320$ & $3.403$ & $3.424$ & $ 3.374$ \\
$18$ & $2.686$ & $3.061$ & $3.147$ & --- & $3.117$ \\
$30$ & $2.240$ & $2.710$ & 2.801 & --- & $2.786$ \\
\hline
\end{tabular}
\label{IV}
\end{table}

\begin{figure}[p]
\caption{
The ground state staggered bond length change as a function of 
bond index from the center of the chain for the $18$ site chain.  
Circles: one  phonon per site, squares: two phonons per site, and 
diamonds: three phonons per site.  The dashed line with crosses is 
the classical approximation.
}
\label{1}
\end{figure}

\begin{figure}[p]
\caption{
The triplet state ($1^3B_u^+$) transition energy as a function of inverse 
chain length.  The same symbols as Fig.\ 1 
(filled diamonds: three phonons per site), with additionally 
$+$: no phonons per site (i.\ e.\ the Pariser-Parr-Pople model).
}
\label{2}
\end{figure}

\begin{figure}[p]
\caption{
The triplet state staggered bond length change as a function of 
bond index from the center of the chain. (The end bonds are not
shown.)  The same symbols as 
Fig.\ 1.  (a) $18$ sites and (b) $30$ sites.
}
\label{3}
\end{figure}

\begin{figure}[p]
\caption{
The $1^1B_u^-$ state staggered bond length change as a 
function of bond index from the center of the chain for the $30$ 
site chain.  The same symbols as Fig.\ 1.
}
\label{4}
\end{figure}


\begin{references}
\bibitem{footnote1}
That is, the limit that the phonon frequency $\rightarrow 0$ and the number
of phonons per site $\rightarrow \infty$.
\bibitem{footnote3}
 At each conjugate gradient 
iteration the super-block target states 
can be re-symmetrised in an efficient 
way provided that the super-block symmetry operators are 
sparse.  This will be so if the primitive block symmetry 
operators are sparse.   Eigenstates of a local density matrix 
 satisfy this property, so the optimised states have sparse 
symmetry operators. However, the `folded-in'
`bare' states have
to be orthogonalised with respect to the optimised states. To 
ensure that these
`bare' states also have sparse symmetry
operators it is necessary to prepare them as eigenstates of a local 
density matrix.  
This is done by preparing a four site bock, 
composed of two end one site blocks (the environment blocks) and 
the middle 
two site repeat unit.
In order to obtain accurate density matrix eigenstates
for the preparation of the `bare' electron-phonon basis
it is necessary to include a number of superblock states of
different symmetries in the density
matrix. Similarly, we use two superblock states 
of
different symmetries for six sites and greater.
\end{references}
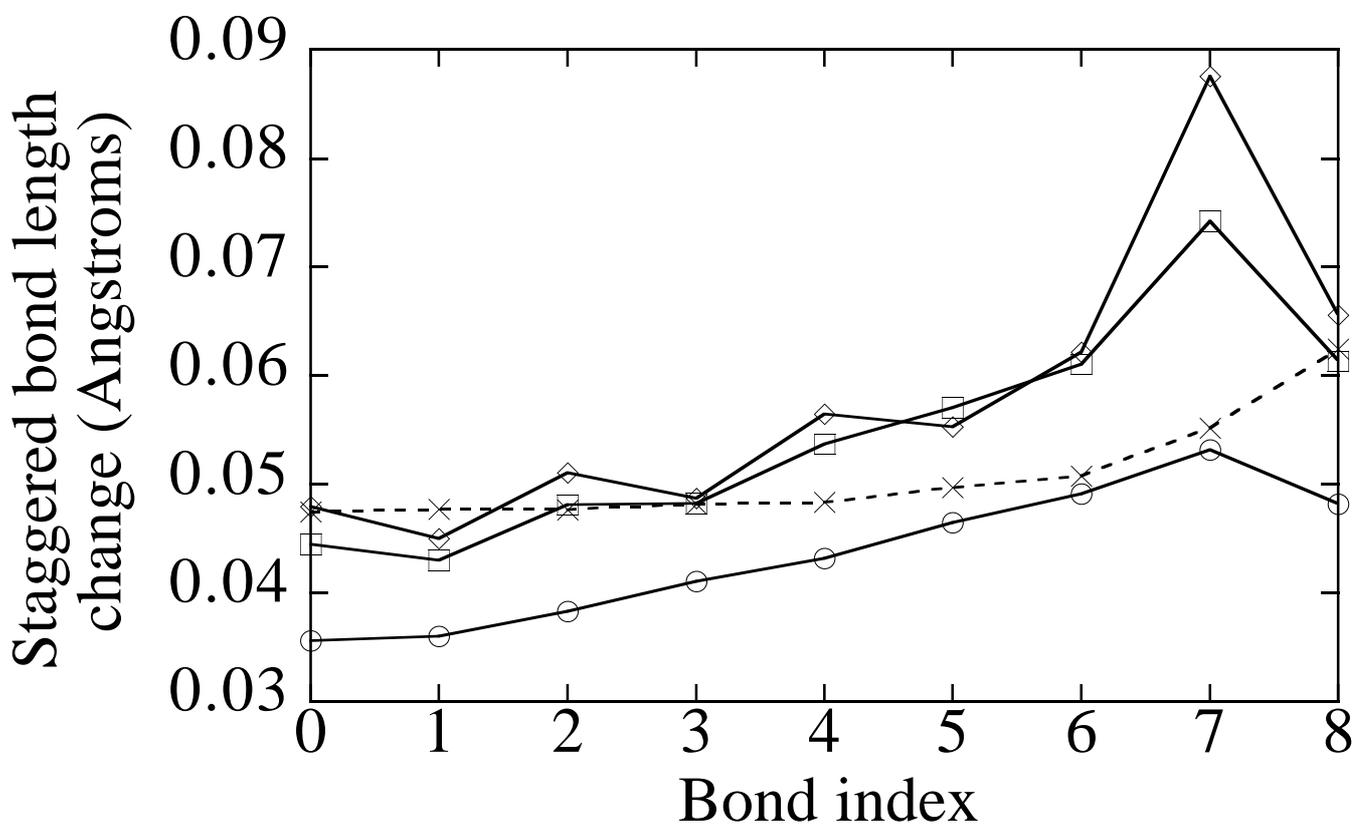
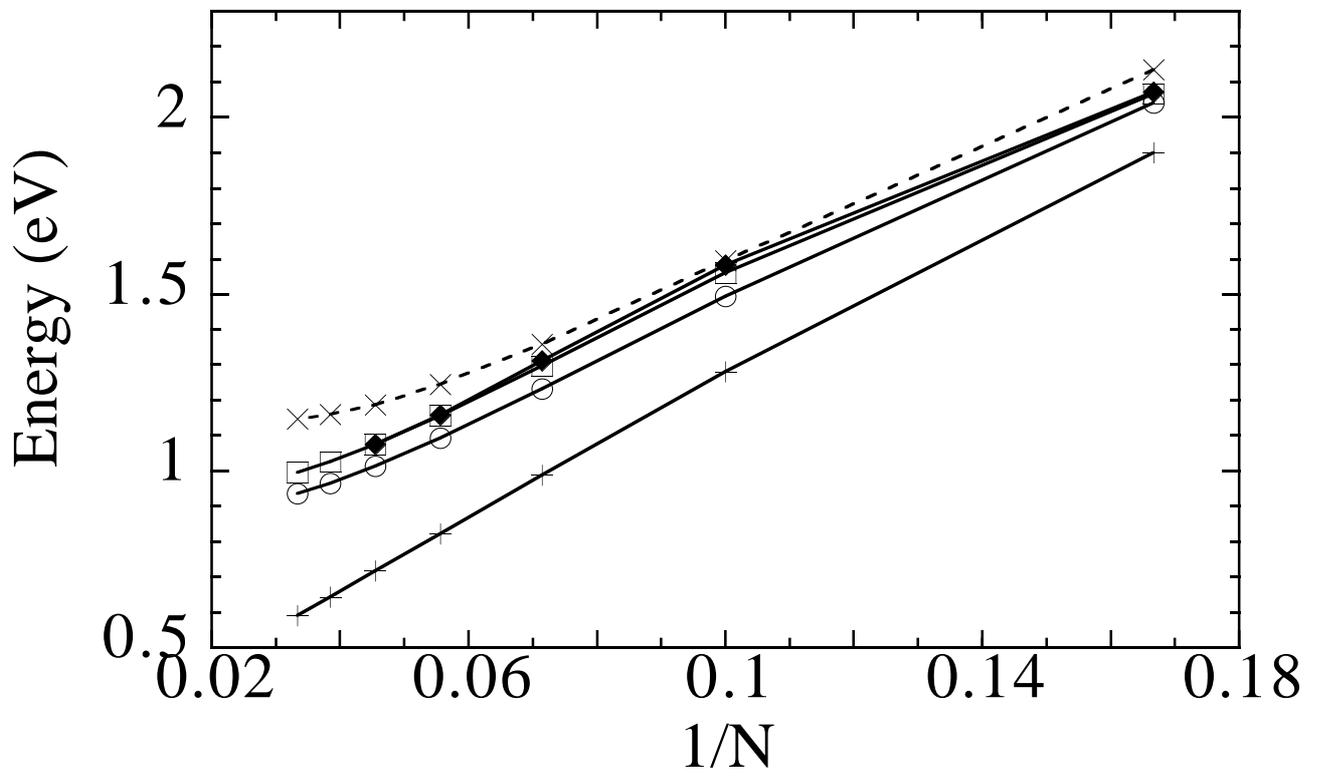
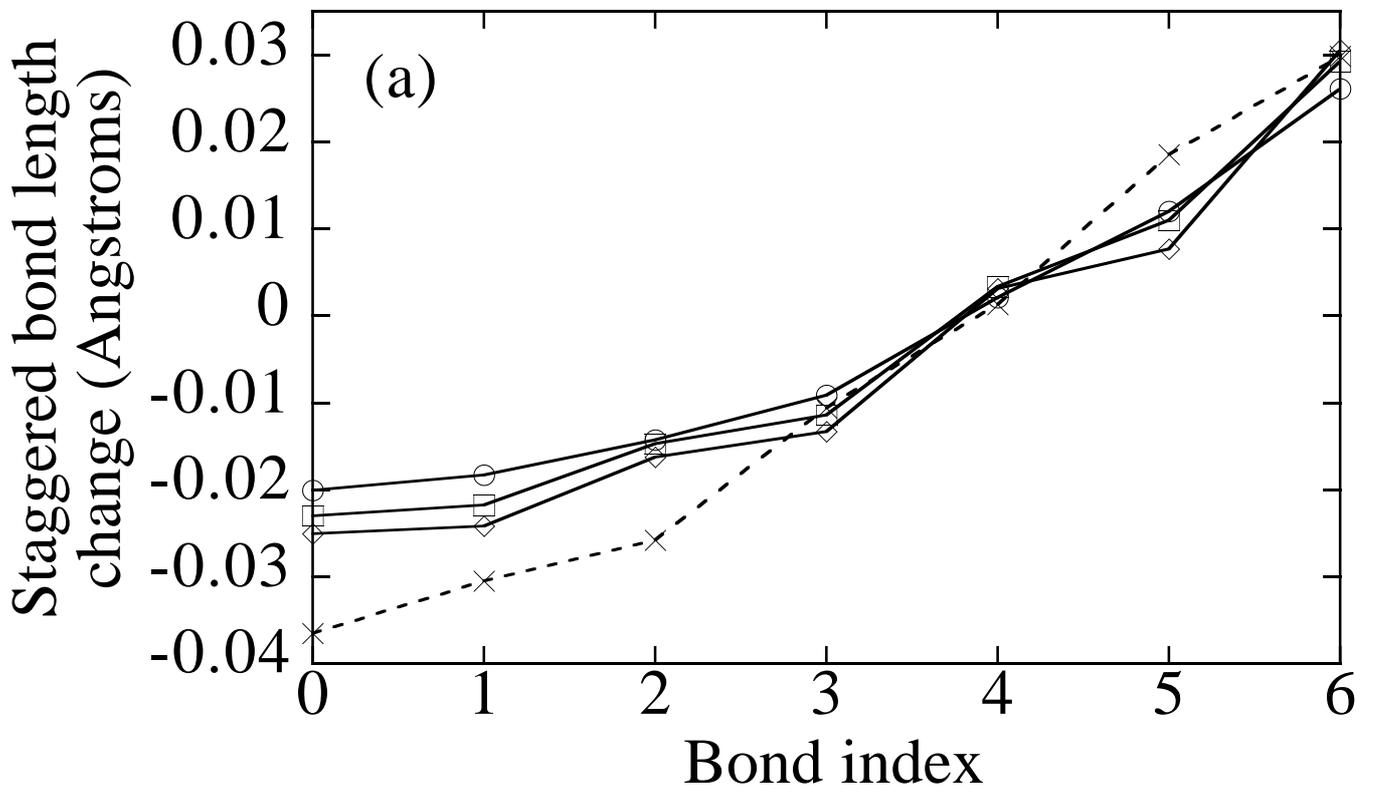
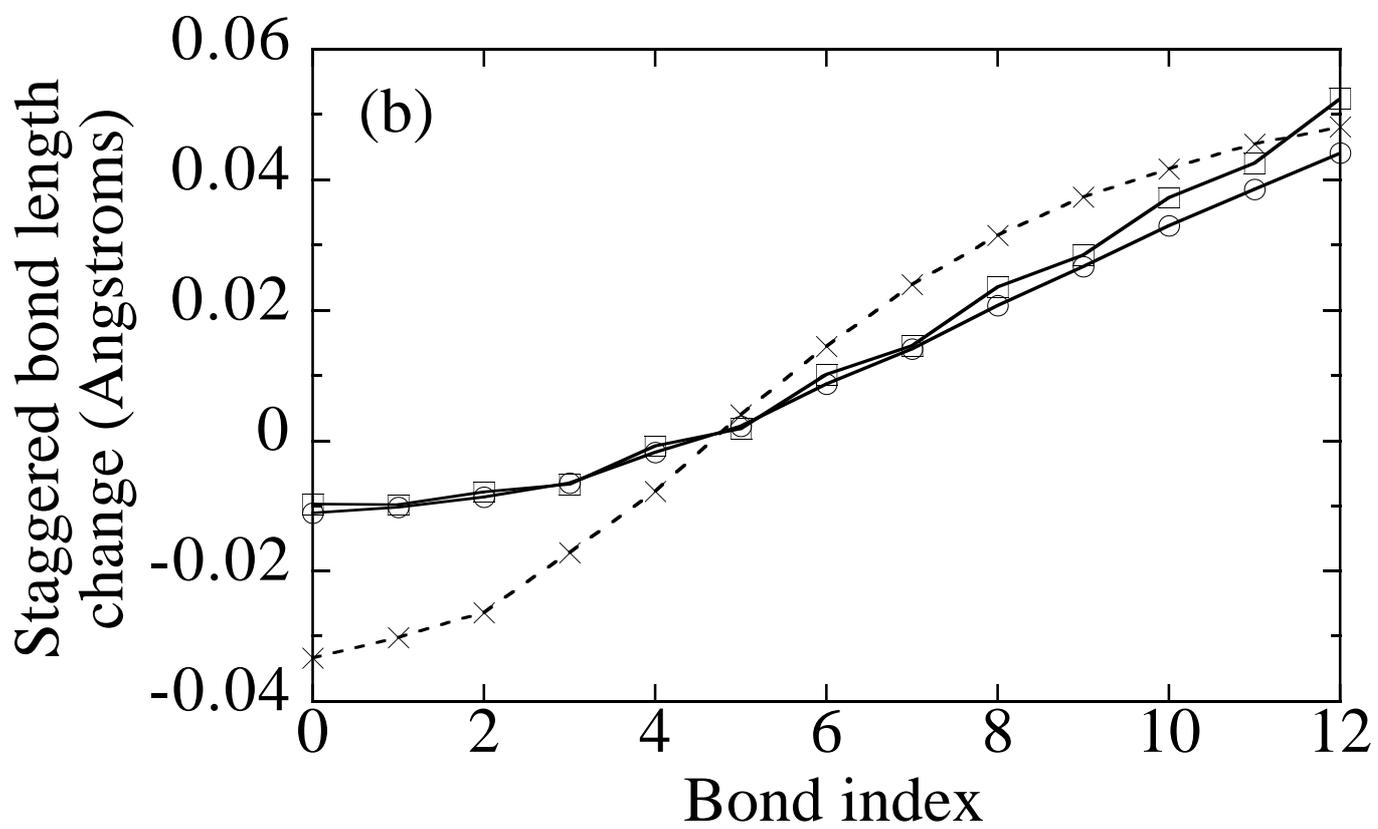
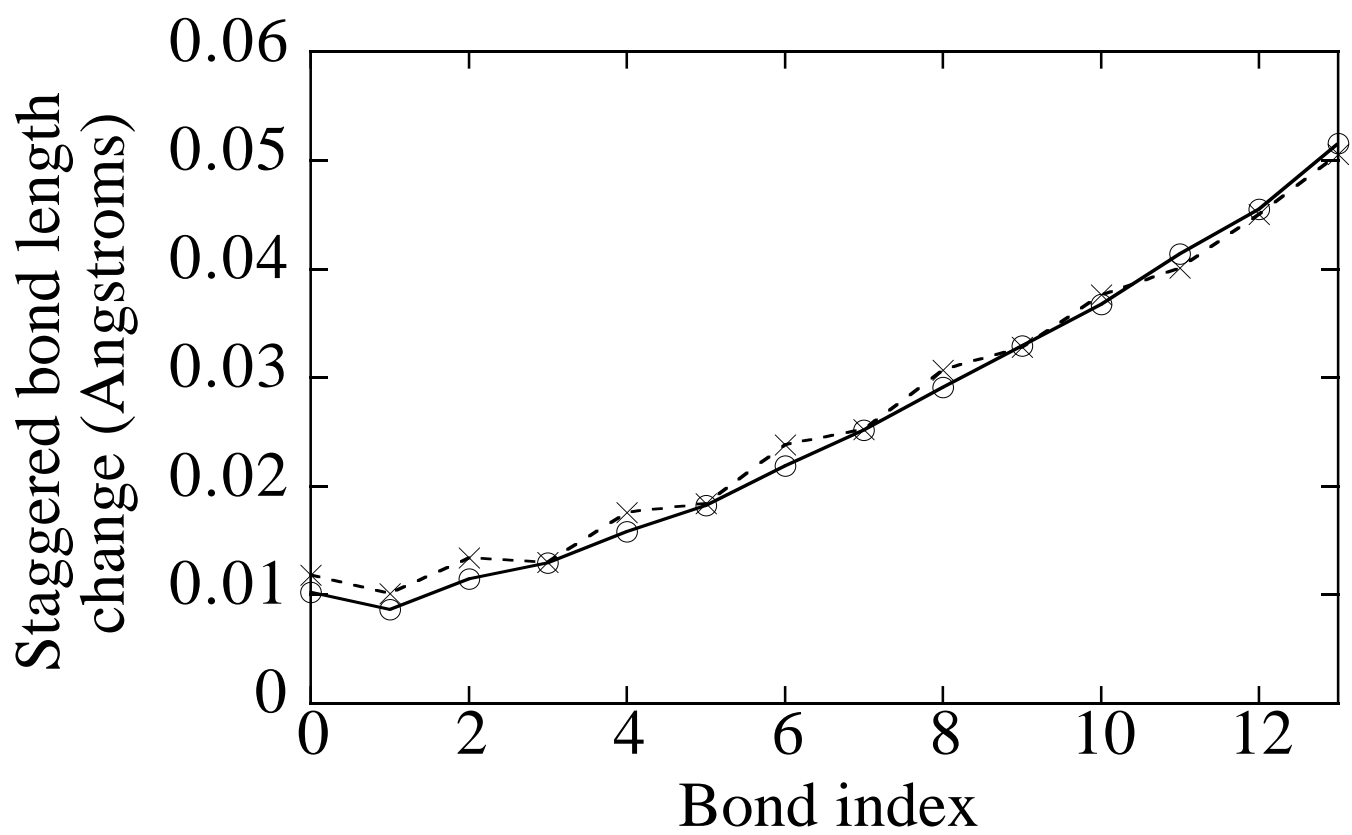
\end{document}